\documentclass[referee]{jfm}

\usepackage{psfrag}
\graphicspath{{./}}
\usepackage{graphicx}
\usepackage{dcolumn}
\usepackage{bm}
\usepackage{natbib}

\title[Inviscid coalescence of drops]{Inviscid coalescence of drops}

\author[L. Duchemin, J. Eggers and C. Josserand]
{L.\ns D\ls U\ls C\ls H\ls E\ls M\ls I\ls N$^1$,\ns 
J.\ns E\ls G\ls G\ls E\ls R\ls S$^2$ \and \break
C.\ns J\ls O\ls S\ls S\ls E\ls R\ls A\ls N\ls D$^3$}

\affiliation{$^1$Department of Mathematics, Imperial College of Science,
 Technology and Medicine, 180 Queen's Gate, London,
  SW7 2BZ, UK\\[\affilskip]
  $^2$Fachbereich Physik, Universit\"at Gesamthochschule Essen,
  45117 Essen, Germany\\[\affilskip]
  $^3$Laboratoire de Mod\'elisation en M\'ecanique, 8,
  rue du Capitaine Scott, 75015 Paris, France}

\begin{document}

\maketitle

\begin{abstract}
We study the coalescence of two drops of an ideal fluid driven by
surface tension. The velocity of approach is taken to
be zero and the dynamical effect of the outer fluid (usually air) 
is neglected. Our approximation is expected to be valid on scales
larger than $\ell_{\nu} = \rho\nu^2/\sigma$, which is $10 nm$ 
for water. 
Using a high-precision boundary integral method, we show that 
the walls of the thin retracting sheet of air between the drops reconnect
in finite time to form a toroidal enclosure. After the initial 
reconnection, retraction starts again, leading to a rapid sequence 
of enclosures. Averaging over the discrete events, we find 
the minimum radius of the liquid bridge connecting the two drops 
to scale like $r_b \propto t^{1/2}$. 
\end{abstract}

\section{Introduction}

Drop coalescence arises in many different contexts, and is crucial 
to our understanding of free surface flows \cite[]{E97}. Examples 
are printing applications \cite[]{CM80,W01}, drop impact
on a fluid surface \cite[]{OP90}, and the coarsening of drop clouds 
and dispersions \cite[]{Metal02,JBKC99,V00}. 
After the two surfaces have merged on
a microscopic scale, surface tension drives an extremely rapid
motion, usually impossible to resolve in either experiment 
\cite[]{BS78,MMNPZ01} or simulation \cite[]{LNSZZ94}. Thus theory
is needed to investigate a possible dependence on initial conditions,
development of small-scale structures during merging, and to
estimate the typical time required for merging. 

A large body of work exists on this problem in the case that viscosity
is dominant and the motion is described by Stokes' equation. 
In the absence of an outer phase this is known as the ``viscous sintering
problem'' \cite[]{F45,H93,MD95}, the inclusion of an outer phase is 
important for many problems governing
the coarsening of dispersion \cite[]{NBG96,V00}. For the two-dimensional 
problem (i.e. for the merging of cylinders) {\it exact} solutions 
exist \cite[]{H90,R92,C02,C03}, which were shown \cite[]{ELS99} 
to be asymptotically equivalent to their three-dimensional counterparts. 
The presence of an outer fluid leads to the formation of a toroidal bubble 
during merging \cite[]{ELS99}, significantly modifying the dynamics. 

\begin{figure}
\begin{center}
\psfrag{omega}{\Large $r_b^2$}
\psfrag{rtip}{\Large $r_b$}
\psfrag{R}{\Large $R$}
\includegraphics[width=0.6\hsize]{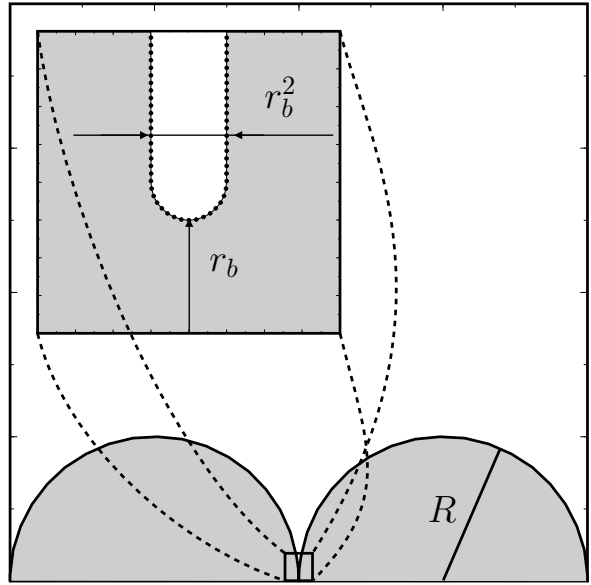}
\caption{\label{init_all} Initial condition. Two drops touching at a 
point are joined by a liquid bridge of radius $r_b$. The inset shows
the width of the gap just above the meniscus to be $w=r_b^2$. The gap's
walls are nearly straight on the scale of $w$. 
                 }
\end{center}
\end{figure}

Fig. \ref{init_all} shows two equal drops of radius $R$ being connected by 
a liquid bridge of radius $r_b$, which is rapidly being pulled 
up by surface tension. The local Reynolds number of this flow
can be estimated as $Re= \sigma r_b/(\rho \nu^2)$, where 
$\sigma$ is the surface tension, $\rho$ the density, and $\nu$
the kinematic viscosity. Thus, regardless of the value of the 
viscosity, the Reynolds number is always small in the initial
phases of the merging, which is equivalent to demanding that 
$r_b \ll \ell_{\nu}$, where $\ell_{\nu} = \nu^2\rho/\sigma$ 
is the viscous length scale. However, $\ell_{\nu}$ is often very 
small (140 \AA  \, for water, and 4 \AA \, for mercury \cite[]{E97}), 
so $r_b \gg \ell_{\nu}$ for a large part of the evolution,
and inviscid theory can be applied. Thus for a wide range
of practical problems the almost inviscid regime, which is the topic
of this letter, is the most relevant. Typically, the viscous
regime will serve as an inner layer that defines the initial condition
for the inviscid problem we are interested in. In general, we
do not have to worry about the initial process of reconnection
\cite[]{ACK01}, which for clean fluids is expected to take place 
over a microscopically small area. 

In the case of a head-on collision of two drops with relative velocity
$V$, considered in \cite[]{OP89}, a purely geometrical consideration
predicts $r_b \approx \sqrt{VRt}$ for two overlapping circles.
The corresponding speed of merging is of the same order as the 
surface-tension-driven merging to be described below, so
$V$ has thus to be taken into account. However, we will restrict ourselves 
here to the case where the velocity of approach is vanishingly small,
a condition that is easily realizable experimentally \cite[]{MMNPZ01}. 
We also do not treat the 
dynamical effect of an outer fluid like air, which might become 
important as the lubrication layer between the approaching drops
becomes very thin \cite[]{ELS99,YD91}. However, this approximation
is consistent with the assumption of a small velocity of approach. 

\section{Initial conditions and scaling laws} 

We consider two identical drops of radius $R$ touching at a point 
where a thin liquid 
bridge of size $r_b$ connects the two drops initially 
(cf. figure \ref{init_all}). The general problem of drops of different 
radii only changes a prefactor in the gap width between the drops
\cite[]{E98}. For the inviscid dynamics considered here, 
all parameters of the problem can be scaled out by writing the
time and space coordinates in units of $\sqrt{\rho R^3/\sigma}$ 
and $R$, respectively. Assuming that the vorticity generated by the
initial viscous motion can be neglected, and using incompressibility, 
the velocity potential
$\varphi$ obeys
\begin{equation}
\Delta \varphi = 0 .
\label{laplace}
\end{equation}
The boundary condition on the free surface amounts to a balance
between surface tension and Bernoulli pressures \cite[]{OP89}:
\begin{equation}
\frac{\partial \varphi}{\partial t} + \frac{1}{2} 
\left( \nabla \varphi\right)^2 - \kappa = 0,
\label{bernoulli}
\end{equation}
where $\kappa$ is the mean curvature of the interface.

We have to solve (\ref{laplace}),(\ref{bernoulli}) with the initial 
condition shown in figure \ref{init_all}, assuming that the 
bridge radius $r_b$ is initially very small (typically $10^{-5}$
in our numerical simulations). Away from the point of contact
at $z=0$, but for $h \ll 1$ the surface has the form $h(z)=(2z)^{1/2}$ and 
$h(z)=(-2z)^{1/2}$ for $z > 0$ and $z < 0$, respectively. 
The width of the gap at a height $r$ is thus 
\begin{equation}
w = r^2 \quad (r_b \ll r \ll 1)
\label{width}
\end{equation}
and since $\partial w/\partial r \ll 1$, the walls are nearly 
parallel. Thus the meniscus, which owing to radial symmetry is 
located along a ring of radius $r_b$, is being pulled straight 
up by a force $2\sigma$ per unit length. 

Assuming that the profile in region (\ref{width}) matches onto 
the bridge on the scale $r\approx r_b$, the curvature at the meniscus can 
be estimated as $\kappa_b \approx r_b^{-2}$, much larger than the 
axial curvature $r_b^{-1}$ of the liquid bridge. Thus, as already 
argued in \cite[]{ELS99}, the axial curvature can be neglected for 
$r_b \ll 1$ and the problem becomes effectively two-dimensional, 
equivalent to the merging of two fluid cylinders. Thus a model 
problem \cite[]{OP89,E98} for the initial motion of the meniscus 
is that of the two dimensional, {\it straight} slot shown in the inset
of figure \ref{init_all}. The eventual widening of the gap can be neglected
on the scale of the gap width $w$. 

\begin{figure}
\begin{center}
\psfrag{x}{$(r - r_b(0))/r_b^2(0)$}
\psfrag{y}{$z / r_b^2(0)$}
\psfrag{rb}{$r_b$}
\psfrag{rc}{$r_c$}
\psfrag{zmax}{$z_{max}$}
\psfrag{pinch_point}{Pinch point}
\includegraphics[width=0.6\hsize]{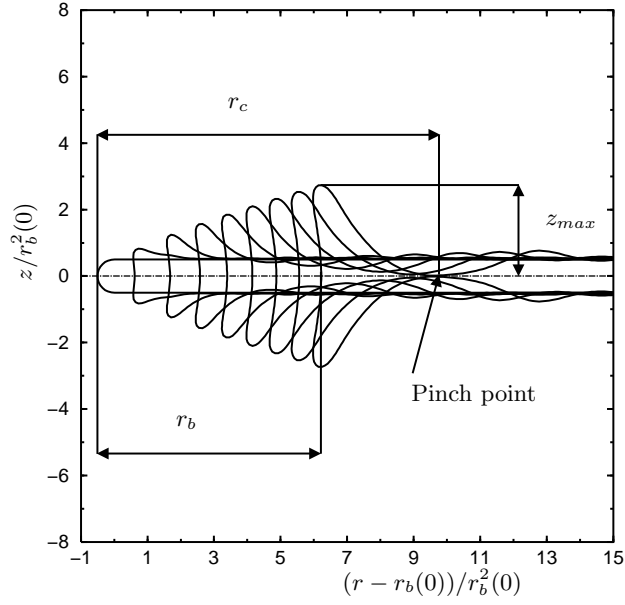}
\caption{\label{figpinch}A sequence of profiles showing the retraction
of the initial meniscus for $r_b(0) = 10^{-5}$. At a time 
$\tau_c=\tau_0 \, w^{3/2} = 7.6 \, w^{3/2}$ the walls of the gap touch and the 
minimum radius $z_{min}$ goes to zero. The distance of this point
from the initial tip of the meniscus is $r_c=r_0 w = 10 \, w$.
     }
\end{center}
\end{figure}

The results of our computations for the full three-dimensional 
problem, to be explained in more detail below, 
are shown in figure \ref{figpinch}. As
the meniscus retracts, the rapid fluid flow past the sides of the
gap creates an under-pressure as described by Bernoulli's equation 
(\ref{bernoulli}), which in turn causes the end to expand into 
a bubble. As the bubble increases in size, capillary waves are
excited in its wake, with amplitude roughly proportional to the
bubble radius. Thus after the amplitude of the capillary wave
has grown to the half width of the slot $w/2$, its two sides touch
and reconnect at a time $\tau_c$. Since the width is the only
length scale in the problem, it follows that the total length $r_c$ the
meniscus has retracted up to the point of reconnection is proportional
to $w$, while the time $\tau_c$ required scales like $w^{3/2}$. 
We thus have 
\begin{equation}
r_c = r_0 w, \quad \tau_c = \tau_0 w^{3/2},
\label{eqpinch}
\end{equation}
where $r_0$, $\tau_0$ are constants to be determined numerically.
Below we find in fact $r_0 = 10, \, \tau_0 = 7.6$. 

After the two sides of the gap have reconnected, this new initial 
condition looks very similar to the original one, except for a non-trivial
velocity field that remains. But since most of the resistance to
the motion {\it before} reconnection is due to the large bubble that was 
left behind, this velocity can be neglected relative to the velocity 
to be generated at the next stage of the motion (more detailed estimates 
are given below). This means that at each step the same motion
repeats itself, but with a slightly larger radius $r_b$. At the 
n-th step we can thus write, analogous to \cite[]{E98}, 
$$
r_b^{n+1} - r_b^n = r_c = r_0(r_b^n)^2, 
$$
and for the times $t_n$ of successive pinching events:
$$
t_{n+1} - t_n = \tau_0 (r_b^n)^3. 
$$

For very small initial $r_b$ reconnection occurs in rapid succession, 
with small relative change of the variables. We can thus write $r_b$ as
a smooth function of $t$, obeying the differential equation
\begin{equation}
\frac{dr_b}{dt} \approx \frac{r_0}{\tau_0}\frac{1}{r_b}
\label{int_pinch}
\end{equation}
which gives, after integration~:
\begin{equation}
r_b \approx \sqrt{\frac{2r_0}{\tau_0}} t^{1/2} .
\label{scaling}
\end{equation}
The scaling law (\ref{scaling}) is the central result of the 
present letter. 
Eventually, when $r_b$ is of the same order than the drop radius,
the widening of the channel overcomes the growth of capillary 
waves, and the enclosure of bubbles stops. This is when the time
scale of retraction $\tau\approx r_b^2\tau_0/(2r_0)$ is shorter
than $\tau_c \approx \tau_0 r_b^3$ characterizing reconnection. 
Thus reconnection will cease when 
$r_b \mbox{\ \raisebox{-.9ex}{$\stackrel{\textstyle >}{\sim}$}\ }1/(2r_0)=
0.05$. We have determined numerically that no more voids are entrapped 
for $r_b>0.035$, in good agreement with our theoretical estimate.
Below we present detailed numerical tests of the scaling predictions,
and investigate further the crucial stage of bubble growth, from which
we are able to extract the numerical constants $r_0,\tau_0$. 

\section{Boundary integral method}

If the flow can be considered potential and incompressible, 
the use of a boundary integral method is advantageous, since the 
velocity field can be calculated from the interface shape. Thus one
only needs to keep track of the interface, represented by a one-dimensional
curve, and grid refinement can be done very efficiently.
The majority of these boundary integral methods require smoothing of 
the surface, in order to avoid short wave length instabilities. 
The method briefly presented here does not require any explicit 
smoothing, except for a redistribution of the points around the 
tip at every time step. 
This redistribution can act as a smoothing, but no damping 
of instabilities, such as an artificial surface viscosity, has been used. 

The dipole formulation used here is very close to the one described by 
Baker, Meiron and Orszag \cite[]{BMO80}, but it needs to be refined to 
be able to resolve the very disparate scales of the drops and of the highly
curved region close to the meniscus. 
At a given time step, we expect the velocity potential $\varphi$ to be 
known, from which we calculate the normal and the tangential velocity 
of the surface. This velocity is then used to advect the surface, and
to advance $\varphi$ using Bernoulli's equation (\ref{bernoulli}). 
The tangential velocity is calculated directly by differentiating 
with respect to the arclength along the interface:
\begin{equation}
u_t  =  \frac{\partial \varphi}{\partial s} 
\label{velot}
\end{equation}
to compute the normal component, we use the vector potential ${\bf A}$
of the velocity field, ${\bf u}=\nabla\times{\bf A}$:
\begin{equation}
u_n  =  \frac{1}{r} \frac{\partial r A_\theta}{\partial s} .
\label{velon}
\end{equation}

Following \cite[]{BMO80}, we first compute the dipole density $\mu$ from 
\begin{equation}
\varphi(M)  =  \mu(M) + \frac{1}{4 \pi} \int_S{(\mu(M) - \mu(M')) 
\frac{\partial}{\partial n} \left( \frac{1}{\lambda} \right)} dS',
\label{green2}
\end{equation}
where $\lambda$ is the distance between points $M$ and $M'$
on the surface. The appearance of $\mu(M)$ in the integrand serves
to subtract the singularity of the normal derivative. Once $\mu$ 
is known, it can be used to calculate the vector potential:
\begin{equation}
{\bf A}(M)  =  \frac{1}{4 \pi} \int_S{(\mu(M') - \mu(M)) \; {\bf n} \times 
\mbox{\boldmath $\nabla$}_s \left( \frac{1}{\lambda}\right)} dS'.
\label{vectpot}
\end{equation}

Classical iterative solutions of (\ref{green2}),(\ref{vectpot}) were 
found to fail for very small bridge radii, so (\ref{green2}),(\ref{vectpot})
were solved by matrix inversion instead. A simple trapezoidal rule was
used to convert the equations into linear systems, which was then 
solved by LU decomposition.
In order to compute the curvature of the surface and the tangential 
derivatives in (\ref{velot}),(\ref{velon}), we re-parametrized the 
integrals by introducing a new integration variable $\zeta$, which equals 
$i$ at grid-point $i$. This avoids instabilities in the cubic spline 
interpolation that would otherwise be present if two points come 
very close together, as it happens at the tip. 

At each time step, the Bernoulli equation and the kinematic condition 
were used to advance the solution using a Crank-Nicolson scheme \cite[]{PTVF}.
The implicit equations were solved by iteration, which required 
less than 10 iterations until a relative error of $10^{-5}$ in
the velocity potential was reached. An explicit Runge-Kutta fourth 
order scheme was also tested, but found to be too unstable for small 
values of $r_b$. 

We also redistribute grid-points at every time step according to the 
their distance from the tip. Cubic splines are used to interpolate 
to the new points. At each time step points are placed on the free 
surface with grid spacing $\delta$; typical values are shown
in figure \ref{pinch}. This spacing is used up to a distance of
$40 \, r_b^2$ from the tip, after which it is gradually increased 
in steps of 2, since much lower resolution is required far from the tip.

\section{Reconnection}
As we have explained above, the retraction of the meniscus is interrupted
by the reconnection of the two sides of the gap, and the distance $r_c$ 
by which the meniscus recoils as well as the time $\tau_c$ required is given
by the scaling relations (\ref{eqpinch}). In figure \ref{figpinch} 
we define typical quantities characterizing the retraction of the meniscus. 
The minimum gap radius $z_{min}$ marks the first trough of a train of 
capillary waves that is generated by the growing bubble. Note that in the 
corresponding simulation in \cite[]{OP90} (cf. figure 4) there is little
or no indication of this growth of capillary waves. We suspect that 
these authors did not follow the retraction for sufficiently long
times, and that the low resolution of their simulation introduced 
additional damping, which smoothed out the capillary waves. 

\begin{figure}
\begin{center}
\psfrag{1_e-11}{$\delta/w = 0.1$}
\psfrag{2_e-11}{$\delta/w = 0.2$}
\psfrag{3_e-11}{$\delta/w = 0.3$}
\psfrag{zmin}{$z_{min} / r_b^2(0)$}
\psfrag{t}{$t / r_b^3(0)$}
\psfrag{Linear approximation}{Linear approximation}
\psfrag{grid spacing}{Grid spacing~:}
\includegraphics[width=0.6\hsize]{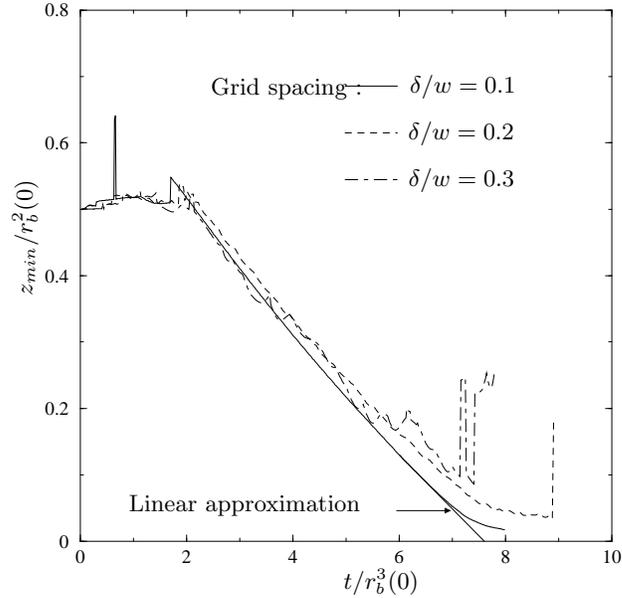}
\caption{The minimum gap radius \label{pinch}$z_{min}/r_b^2(0)$ 
plotted against $t / r_b^3(0)$. 
The initial value of $r_b$ is $10^{-5}$ and the three 
resolutions correspond 
to the minimum distance between points in the tip region. 
The linear extrapolation gives $t_{pinch} \simeq 7.6 \; r_b^3(0)$}
\end{center}
\end{figure}

As seen in figure \ref{pinch}, the time dependence of the 
minimum gap radius $z_{min}$ converges towards a close to linear 
behavior as the resolution is increased. Extrapolation towards 
$z_{min}=0$ thus gives a reliable estimate of the time required for 
reconnection. Although the walls of the gap do not interact physically,
errors of our boundary integral description grow large as two surfaces 
become close to each other. The reason is that the distance $\lambda$
between points varies on scale $z_{min}$ close to the minimum, so 
the grid spacing $\delta$ always needs to be smaller than $z_{min}$. 

\begin{figure*}
\begin{center}
\psfrag{0}{$0$}
\psfrag{-0.002}{$-0.002$}
\psfrag{0.002}{$0.002$}
\includegraphics[height=\hsize, angle=-90]{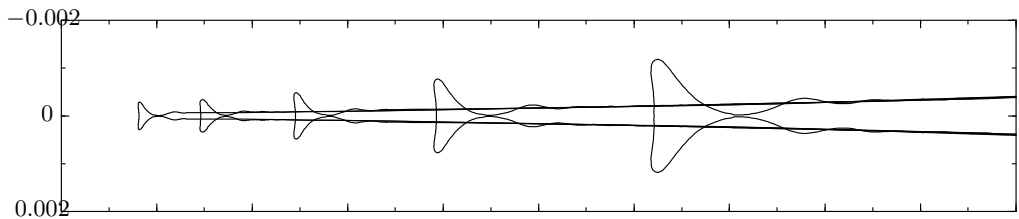}
\caption{\label{fig:voids}Successive entrapment of voids during the 
coalescence for an initial liquid bridge radius of $r_b = 0.008$.
After every reconnection, the void is extracted from the profile and a new 
computation begins, with a null initial velocity field.}
\end{center}
\end{figure*}

From the simulations we deduce the values $r_0 = 10$ and $\tau_0 = 7.6$
for the reduced retraction length and time already reported in 
section 2. Here the underlying assumption is that the dynamics is
controlled by the local gap width alone. To test this idea, 
we have computed a sequence of pinch events as shown in figure 
\ref{fig:voids}. When $z_{min}$ has gone down to about 10 \% 
of the local gap radius $w/2$, the gap is cut at about $w/2$
behind the minimum and new points are introduced along the
new surface. Our method of redistributing points automatically 
introduced a certain smoothing, which was enough for the simulation
to continue. Obtaining a new initial 
condition for the velocity profile proved to be much more difficult. 
Simply extrapolating the velocity potential $\varphi$ before the 
surgery to the new initial condition led to instabilities that 
could no longer be controlled numerically, so instead we had to put
the velocity field to zero. This is justified by the fact that the 
gap position very quickly re-assumes its retraction velocity after the bubble 
is left behind, as we discuss in more detail below. 

\begin{figure}
\begin{center}
\psfrag{results}{$r_c/\tau_c = f(r_b)$}
\psfrag{scaling}{$(r_0/\tau_0)/r_b$}
\psfrag{rb_t}{$r_b(t)$}
\psfrag{sqrt}{$f(t) = \sqrt{2 r_0 t / \tau_0}$}
\psfrag{t}{$t$}
\psfrag{rb}{$r_b(t)$}
\includegraphics[width=0.6\hsize]{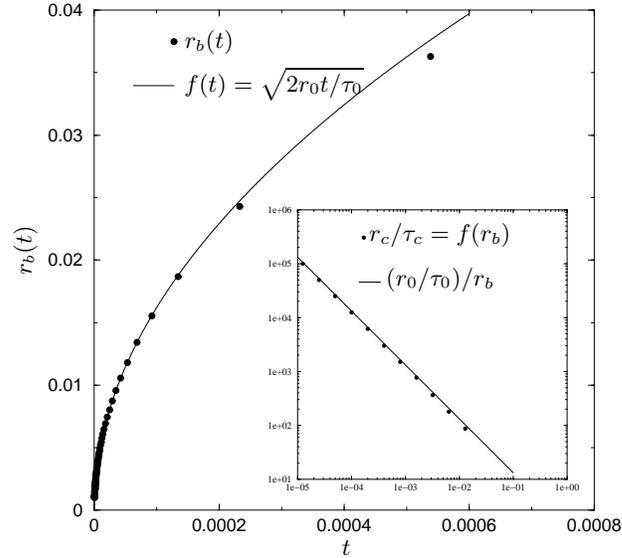}
\caption{\label{fig:rb_all}The minimum radius $r_b$ as a function
of time (dots), compared to the theoretical prediction 
$\sqrt{2r_0t/\tau_0}$ (full line). Inset: the ratio $r_c/\tau_c$ 
as a function of the initial 
radius $r_b^n$ varying between $1.25\cdot 10^{-5}$ and $1.28\cdot 10^{-2}$. 
The time for pinching $\tau_c$ was approximated using a linear 
extrapolation of $z_{min}$. The numerical results show very good agreement 
with the expected scaling law.
    }
\end{center}
\end{figure}

As illustrated in figure \ref{fig:voids}, this leads to a self-similar
succession of pinch-off events. Each simulation was started from a
new value of the bridge radius $r_b^n$. The typical gap width at
the meniscus is then $w=(r_b^n)^2$. A more quantitative test of the scalings
employed in section 2 is presented in figure \ref{fig:rb_all}, where 
we plot the bridge radius $r_b$ as a function of time and, in the inset, 
$r_c/\tau_c=(r_0/\tau_0)/r_b^n$ as function of the bridge 
radius at the time of pinching. The excellent agreement with the 
predicted scaling behavior confirms our assumption that the local 
dynamics only depends on the gap width at the corresponding radius
$r_b^n$.

We also did not follow the evolution of the bubble after it was cut 
off from the gap. Since it starts from a highly non-circular shape,
it is expected to perform large amplitude oscillations. Remembering
that the bubble is really a torus in three-dimensional space, 
it will also be unstable with respect to the Rayleigh instability
\cite[]{DR} and break up into a sequence of smaller bubbles. Evidently,
this instability breaks the rotational symmetry and is thus well 
beyond the scope of the present work. 

\section{Dynamics of retraction}
We now study the individual retraction events, characterized by a 
mass of fluid being accelerated by two line forces, in greater detail.
Thus if 
$$ \frac{d r_b}{dt}=v_{tip} $$
is the velocity of the receding tip, the force balance reads
\begin{equation}
\frac{d}{dt}\left(M_{tip}\frac{dr_b}{dt} \right) = 2,
\label{momentum}
\end{equation}
where $M_{tip}$ is the mass being accelerated. This ``added mass'' is 
being pushed along by the structure of maximum radius $z_{max}$ 
that is forming at the end of the gap, and thus 
$M_{tip}\approx C z_{max}^2$ \cite[]{LL}, where $C$ is a numerical 
constant coming from the geometry of the void profile. Hence the equation
of motion becomes 
\begin{equation}
\frac{d}{dt}\left(C z_{max}^2\frac{dr_b}{dt} \right) = 2.
\label{momentum2}
\end{equation}

\begin{figure*}
\begin{center}
\psfrag{x}{$t/r_b^3(0)$}
\psfrag{y}{$\delta r/r_b^2(0)$}
\begin{minipage}[c]{.46\linewidth}
\includegraphics[width=0.9\hsize]{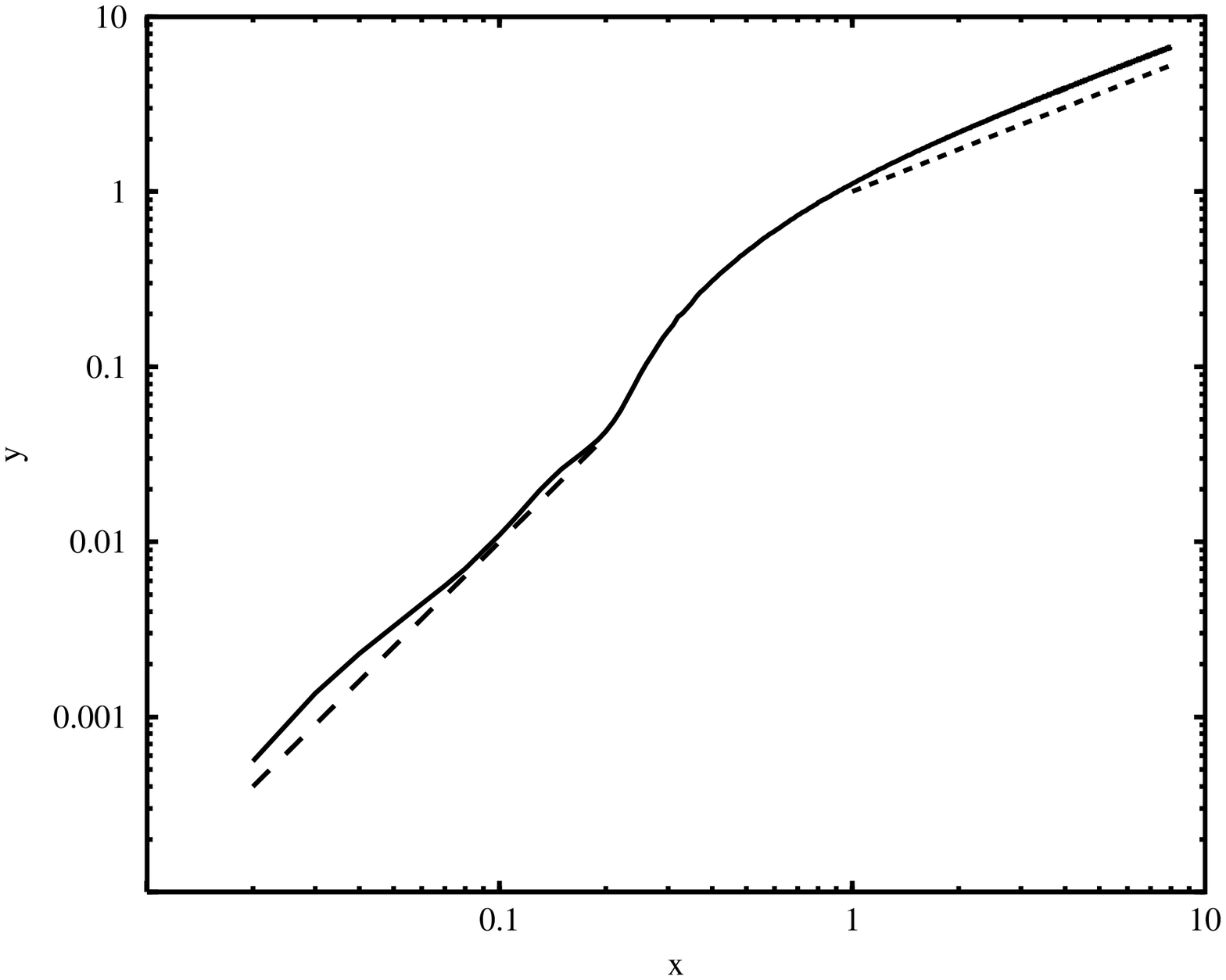}
\end{minipage} \hfill
\begin{minipage}[c]{.46\linewidth}
\psfrag{x}{$t/r_b^3(0)$}
\psfrag{y}{$z_{max}(t)/r_b^2(0)$}
\includegraphics[width=0.9\hsize]{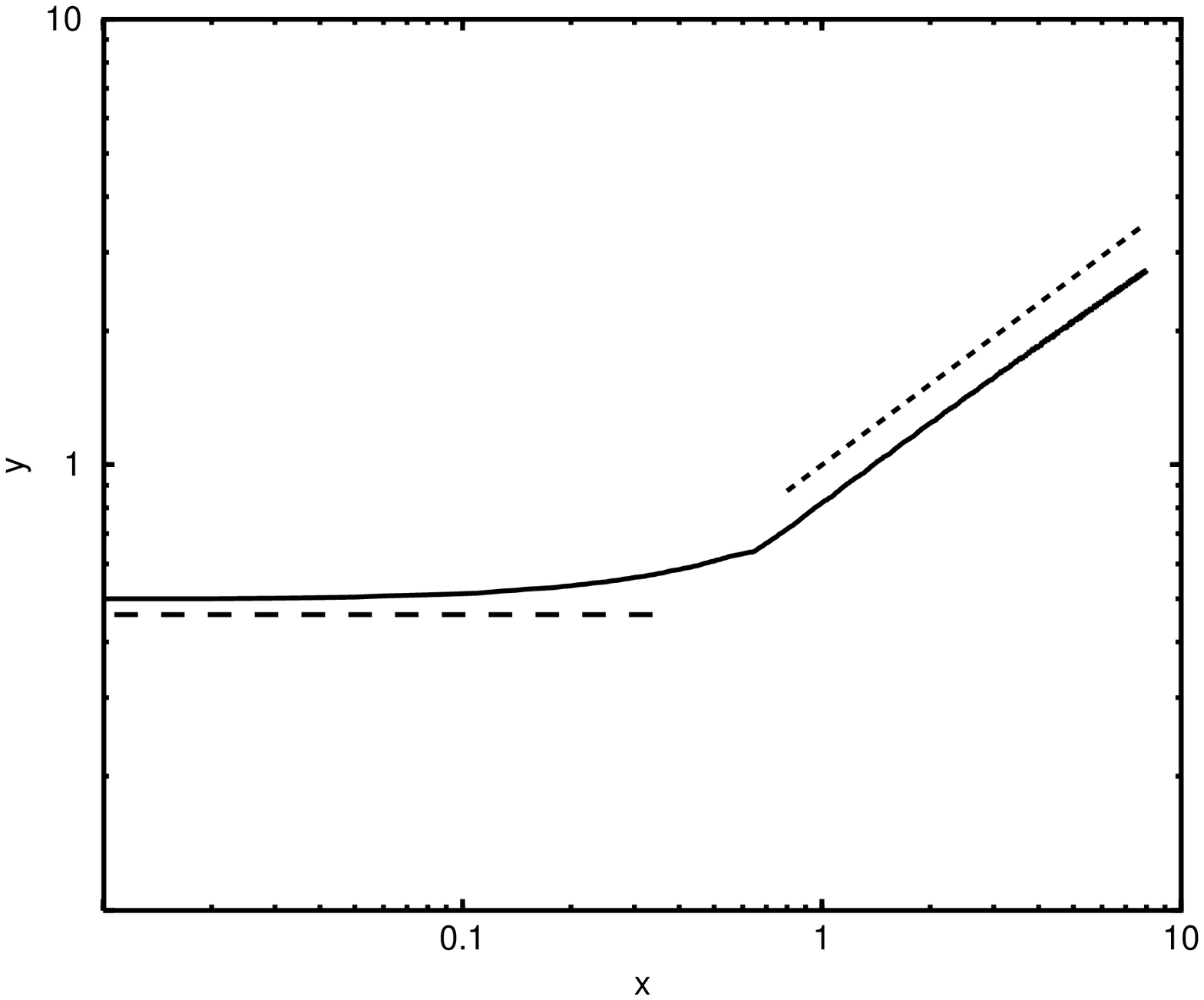}
\end{minipage}
\caption{\label{rtip_zmax}
Two quantities characterizing retraction, 
$\delta r = r_b(t) - r_b(0)$ and $z_{max}$ 
as functions of time in rescaled units. Long-dashed and dotted lines 
represent power-law approximations to the early and long-time behavior,
respectively. We find $\delta r\propto t^2$ for early times, 
while $z_{max}$ remains constant. For late times 
$\delta r\propto t^{0.8}$ and $z_{max}\propto t^{0.6}$; both behaviors
are in agreement with (\ref{momentum2}).
          }
\end{center}
\end{figure*}

For short times, the bubble does not have time to grow, so 
$z_{max}$ is approximately constant and given by the initial 
gap radius:$z_{max} \approx r_b^2(0)/2$. This corresponds to a
constant mass being accelerated by a constant force, and (\ref{momentum2})
leads to a quadratic growth of the retraction distance 
$\delta r_b(t) = r_b(t)-r_b(0)\propto t^2$. This is confirmed 
by the early time behavior of $\delta r_b(t)$ as shown in 
figure \ref{rtip_zmax}. Note that, consistent with (\ref{momentum2}),
$z_{max}$ remains constant. 

\begin{figure*}
\begin{center}
\psfrag{x}{$t/r_b^3(0)$}
\psfrag{y}{$v_{tip} r_b(0)$}
\includegraphics[width=0.6\hsize]{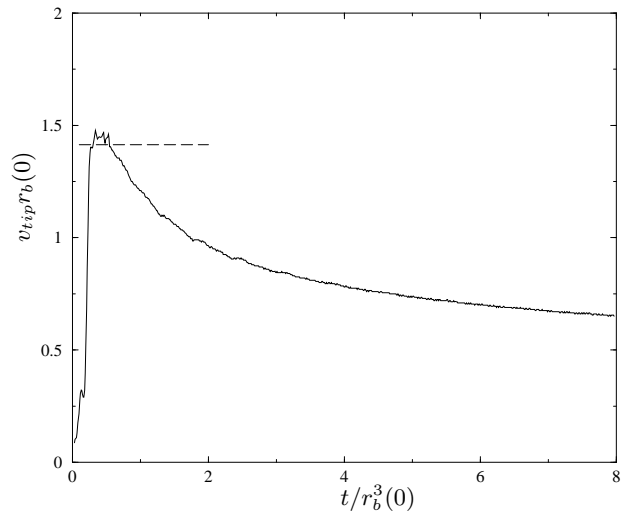}
\caption{The speed of the retracting bridge 
$v_{tip}=dr_b(t) / dt$
as a function of time in rescaled units. The Culick-Taylor velocity
($\sqrt{2}$ in these units) is represented by the dashed 
segment.}
\label{vitrb} 
\end{center}
\end{figure*}

After this initial period of acceleration, the bubble radius 
$z_{max}$ starts to grow and the speed of retraction $v_{tip}$ reaches a
maximum, as seen in figure \ref{vitrb}. This maximum must be 
set by the initial width $w$ of the gap, and thus dimensional arguments 
lead to 
\begin{equation}
v_c \approx \sqrt{2/w}.
\label{Culick}
\end{equation}
The prefactor in (\ref{Culick}) comes from balancing the inertial
term $v_c^2/2$ with the surface tension force $\kappa$ in (\ref{bernoulli}),
in analogy with the arguments of Culick and Taylor \cite[]{CuTa1,CuTa2} for
receding soap films. The curvature $\kappa$ has been approximated by 
$1/w$. As confirmed by figure \ref{vitrb}, the maximum of $v_{tip}$
is well approximated by the estimate (\ref{Culick}). 

After reaching a maximum, the speed of retraction decreases steadily,
as the bubble grows and with it the added mass that has to be dragged
along. The transversal bubble expansion is due to the rapid fluid 
motion along its sides which, according to Bernoulli's equation 
(\ref{bernoulli}), causes an under-pressure. Conversely, at the stagnation
point behind the bubble the pressure is high and the bubble is curved
inward (cf. figure \ref{figpinch}). 
We do not yet have a fully quantitative theory of the bubble
expansion, since this would require a precise knowledge of the bubble's 
shape. Namely, the fluid speed $v_m$ past the crest of the 
bubble is determined by its curvature $\kappa_c$ \cite[]{L93}: 
$v_{m}=v_{tip}\kappa_c z_{max}$, in analogy to 
the flow past an ellipsoidal body. To close the system of equations,
we would need an expression for $\kappa_c$. However, we notice from figure
\ref{rtip_zmax} that the temporal growth of the bubble size $z_{max}$
is well described by a power law: $z_{max} \propto t^{0.6}$.
Plugging this into equation (\ref{momentum2}) we find 
\begin{equation}
\delta r_b \propto t^{0.8},
\label{power}
\end{equation}
in good agreement with simulations, cf. figure \ref{rtip_zmax}. 
The range of validity of the power laws proposed here can of course
never exceed an order of magnitude, since the gap pinches off after
time $t\approx 10 r_b^3$. 

Eventually, when the toroidal bubble separates from the gap,
the velocity $v_{tip}$ has decreased to about half of $v_c$. 
Therefore, the effect of the dynamical pressure $v_{tip}^2/2$
is reduced considerably relative to the capillary pressure. 
Numerically, we find that the capillary force is at least $4$ 
times bigger than the dynamical pressure, which indicates that
the velocity field can safely be neglected at reconnection,
as we are forced to do owing to limitations of our numerical 
technique. 

\section{Discussion}

We have shown that the merging of low viscosity fluid droplets leads
to a self-similar sequence of void entrapments. It is interesting
to note that the same power law behavior (\ref{scaling}) of $r_b$
can be formally derived from a continuous evolution if $v_{tip}$ is 
assumed to be the Culick velocity (\ref{Culick}). If the 
gap width $w$ is estimated form the geometrical constraint 
$w\approx r_b^2$, this immediately leads to 
$\partial_t r_b \approx \sqrt{2}/r_b$, which can be integrated to give
a power $t^{1/2}$. This is the argument given in \cite[]{ELS99},
which did not take reconnection into account. The reason it ends
up to give the correct answer (apart from the prefactor) is that 
the size of the gap tip is 
rescaled to agree with the geometrical estimate (\ref{width}) at
each reconnection event. Thus although the bubble actually grows
to a much larger size than $r_b^2$, the balance implied by the 
above argument is actually true {\it on average}. 

It might be equally tempting \cite[]{L00} to apply the same reasoning
to the force balance (\ref{momentum}), by approximating (at least on
average) the added mass by $M_{tip}\approx Cz_{max}^2\propto r_b^4$. 
Integrating the corresponding equation of motion leads to 
$r_b\propto t^{2/5}$. This apparent paradox is explained by the fact
that the reconnection events destroy the momentum conservation
implied by (\ref{momentum}). Owing to bubble growth, momentum 
is distributed over a much larger volume than estimated from the
simple geometrical argument. Accordingly, in the asymptotic limit
of $t\ll1$ one obtains a motion that is {\it faster} than that 
given by the full calculation including reconnection. 

We would finally like to point out some questions inspired by this work.
Firstly, it would be nice to develop a more complete theory of the bubble
growth at the end of the receding meniscus. Secondly, we are not yet able
to fully treat the velocity field after reconnection. Such a treatment
may lead to an increase in fluctuations and perhaps some randomness
during retraction. As pointed out in \cite[]{OP90}, a finite velocity 
of approach will increase the likelihood of bubble entrapment during
coalescence. Other interesting generalizations not yet considered in 
the present paper are the effect of an external fluid as well as
viscous corrections. Clearly, a number of theoretical questions remain
open. Perhaps more importantly, detailed experimental studies are
called for, for example to verify the phenomenon of bubble entrainment
predicted by our analysis. 

\begin{acknowledgments}
It is our pleasure to thank St\'ephane Zaleski for its constant encouragement
during this work.

\end{acknowledgments}

\bibliography{fusion}
\bibliographystyle{jfm}

\end{document}